\documentclass[sigconf]{acmart}

\renewcommand\footnotetextcopyrightpermission[1]{} % Removes footnote with conference info
\usepackage{graphicx}
\usepackage{listings}
\usepackage{xcolor}

\begin{document}

% Critical Inker: Analyze arguments to improve critical thinking
\title{Critical Inker: Scaffolding Critical Thinking in AI-Assisted Writing Through Socratic Questioning}

% Authors (Anonymized for review if needed, or add your details)
\author{Philipp Hugenroth}
\affiliation{%
  \institution{MIT Media Lab}
  \city{Cambridge}
  \country{United States}
}
\email{philhuge@media.mit.edu}

\author{Valdemar Danry}
\affiliation{%
  \institution{MIT Media Lab}
  \city{Cambridge}
  \country{United States}
}
\email{vdanry@mit.edu}

\author{Pattie Maes}
\affiliation{%
  \institution{MIT Media Lab}
  \city{Cambridge}
  \country{United States}
}
\email{pattie@media.mit.edu}

\begin{abstract}
As Large Language Models (LLMs) increasingly automate writing tasks, there is a growing risk of cognitive deskilling where users offload critical thinking to the system. To address this, we introduce \textit{Critical Inker}, a writing tool designed to scaffold critical reflection during writing through logical analysis and socratic feedback. We present two methods: (1) A Socratic chatbot using questions to help them realize and fix logical errors in their writing and (2) Visual Feedback, which highlights logical errors in the text without dialog. We detail the technical implementation of the system and evaluate its argument extraction and logical validity accuracy. Our evaluation shows a 91.2\% argument overlap with ground truth argument annotations and 87\% validity accuracy. Finally, we conducted a small-scale pilot and discuss early qualitative results.
\end{abstract}

% CCS Concepts
\begin{CCSXML}
<ccs2012>
   <concept>
       <concept_id>10003120.10003121</concept_id>
       <concept_desc>Human-centered computing~Human computer interaction (HCI)</concept_desc>
       <concept_significance>500</concept_significance>
   </concept>
   <concept>
       <concept_id>10003120.10003121.10003124.10010865</concept_id>
       <concept_desc>Human-centered computing~Interactive systems and tools</concept_desc>
       <concept_significance>300</concept_significance>
   </concept>
</ccs2012>
\end{CCSXML}

% \acmConference[CHI '26]{CHI Conference on Human Factors in Computing Systems}{April 26--May 1, 2026}{Barcelona, Spain}
\ccsdesc[500]{Human-centered computing~Human computer interaction (HCI)}
\ccsdesc[300]{Human-centered computing~Interactive systems and tools}

\keywords{AI-assisted writing, cognitive deskilling, Socratic method, productive friction, critical thinking}

\begin{teaserfigure}
  \includegraphics[width=\textwidth]{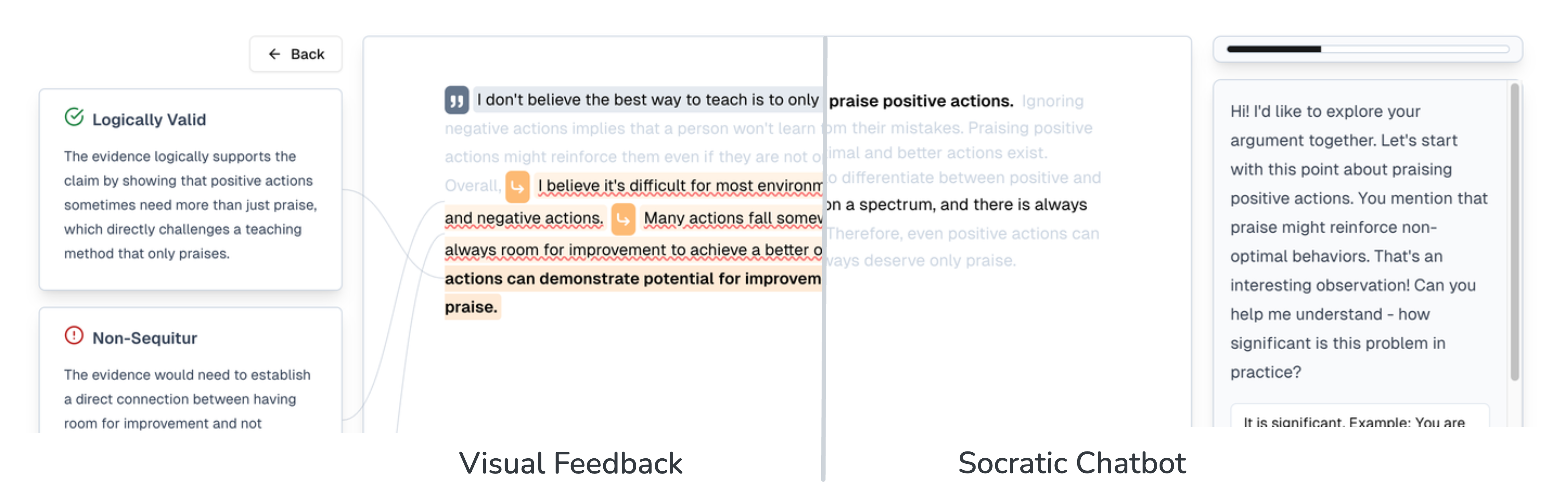}
  \caption{Methods for critical reflection: Visual feedback (Left) shows the argument structure directly with validity checking. The Socratic Chatbot (Right) engages the user in a dialogue, focusing on one segment at a time.}
  \label{fig:teaser}
\end{teaserfigure}

\maketitle

% -------------------------------------------------------------------------
% 1. INTRODUCTION
% -------------------------------------------------------------------------

\section{Introduction}

The adoption of LLMs in writing workflows offers significant efficiency gains, but raises concerns about cognitive offloading and weakening of critical thinking. Writing is deeply intertwined with thinking, an iterative cognitive process involving planning, evaluation, and revision through which reasoning is constructed and refined \cite{flower1981cognitive}. Externalizing thoughts through mediums such as writing enables individuals to identify logical gaps, reconsider assumptions, and develop coherent arguments \cite{pea1987cognitive}, making writing essential in law\cite{swisher1987introduction}, science\cite{writingthinking2025}, journalism\cite{porlezza2019accuracy}, and public safety\cite{redwine2003importance}. Argumentative writing particularly requires writers to articulate claims, justify them with evidence, and ensure logical coherence, making it a critical vehicle for developing reasoning and critical thinking skills \cite{stab2017recognizing}.

LLM-powered writing assistants have been rapidly integrated into educational, professional, and creative environments \cite{lee2022coauthor, yuan2022wordcraft, zhang2023visar}, improving productivity and reducing effort \cite{firaina2023exploring}. However, this can shift writing from an active reasoning process to supervising AI-generated output.

Research suggests this has unintended cognitive consequences. Studies show AI assistants lead users to offload reasoning without fully engaging underlying cognitive processes \cite{buccinca2021trust, fan2025beware}. While AI improves immediate performance, it may not build lasting reasoning ability---dialogues with AI reduce belief in misinformation during interaction but don't lead to durable discernment skills \cite{rani2025dialogues}. Neuroscience evidence suggests heavy reliance on AI writing reduces engagement of brain regions associated with deep cognitive processing \cite{kosmyna2025your}, potentially leading to deskilling \cite{budzyn2025endoscopist}. Co-writing with language models systematically shifts users' views toward model-favored positions and homogenizes content \cite{jakesch2023co, agarwal2025ai}, raising concerns that efficiency-focused assistants may inadvertently weaken engagement in reasoning processes and lead to unfavorable outcomes.

Most existing AI writing assistants focus on producing high-quality text with minimal effort. Although these systems are effective in improving the final product, they reduce opportunities for reflection and critical evaluation by providing direct answers and revisions. In contrast, interaction designs that encourage active reasoning, such as cognitive forcing functions or metacognitive interventions, improve users' ability to critically evaluate information \cite{buccinca2021trust, tankelevitch2024metacognitive, zhang2025friction, weber2024, wambsganss2024, seyed2025}. However, these designs are often described as more difficult and less preferred.

A promising alternative frames AI interventions as questions guiding users to conclusions themselves, drawing from the Socratic method. Prior work shows reframing AI explanations as questions improves users' ability to identify flawed reasoning and increases logical discernment \cite{danry2023don}. Question-driven guidance can support critical thinking and reflection \cite{le2019technology, xi2025investigating}. However, applying Socratic interaction to writing requires understanding the underlying logical structure of arguments.

Recent advances in LLMs have enabled the extraction of informal logical structures from natural language \cite{zhang2023visar,miandoab2025intelliproof}, evaluation of logical validity and the detecting fallacious reasoning \cite{ruiz2023detecting,jeong2025large}. By combining argument-structured representations with Socratic interaction, AI systems can enable targeted interventions that guide users toward independently recognizing and resolving logical issues.

Despite progress in AI-assisted writing and reflective interaction design, existing writing assistants rarely integrate argument-structured analysis with Socratic interaction to scaffold reasoning during writing. To address this gap, we introduce \textit{Critical Inker}, a writing tool designed to scaffold critical reflection during writing through AI-assisted argument analysis and Socratic interaction. Critical Inker identifies claims, supporting evidence, and potential logical flaws in argumentative essays, and guides users through structured questioning that encourages them to articulate and resolve issues themselves. By combining LLM-based logical analysis with interaction designs that preserve cognitive engagement, Critical Inker demonstrates how AI writing assistants can support reasoning rather than replace it.

\section{System Design}
Critical Inker is a web-based writing tool that provides argument-aware feedback on argumentative essays through two interaction modalities: \textit{Visual Feedback} and \textit{Socratic Chatbot}. Both modes share the same underlying argument analysis but differ in how they surface issues and engage users in reflection.

\subsection{Core Workflow}

\begin{figure*}[h]
  \centering
  \includegraphics[width=\linewidth]{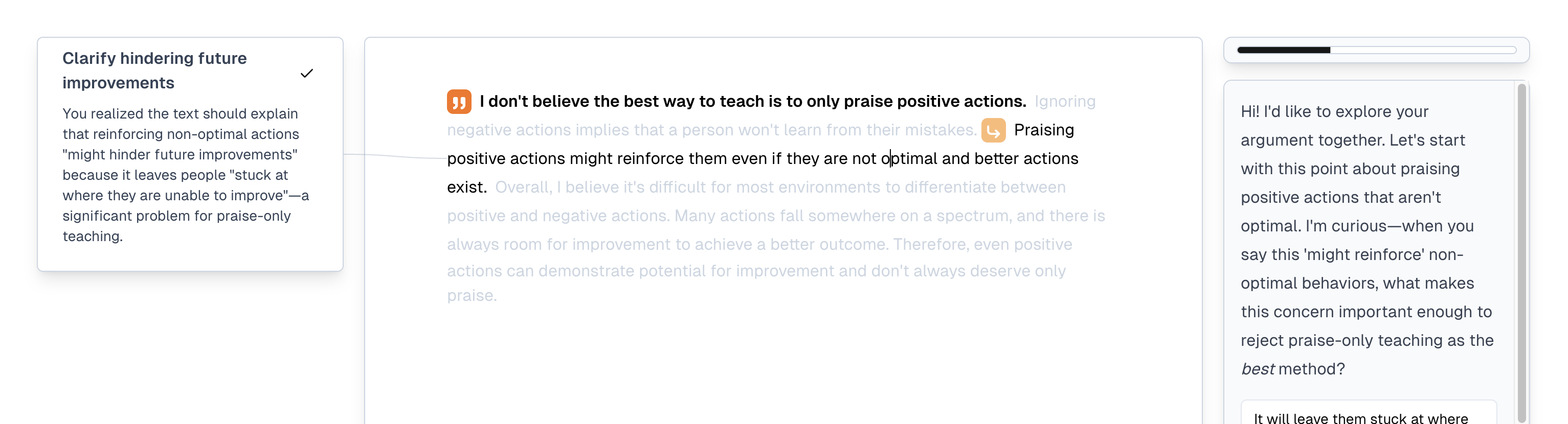}
  \caption{Socratic Chatbot: On successful Socratic scaffolding, the system converts the user's verbalized intention into an actionable comment marker anchored to the text.}
  \label{fig:ui}
\end{figure*}

The interaction follows three stages:

\begin{itemize}
\item{\textbf{1. Write.}} Users write argumentative essays in a minimal text editor without real-time feedback. When ready for analysis, they trigger the system to evaluate their argumentation.

\item{\textbf{2. Analyze.}} The system extracts the argument structure using a LLM pipeline (detailed in Section \ref{sec:prompting}): first identifying claims, premises, and their relationships, then validating logical connections.

\item{\textbf{3. Reflect.}} The two modalities diverge here:
\begin{itemize}
    \item \textit{Visual Feedback}:
    The main claim is highlighted and the full logical structure can be explored. Logical flaws are redlined with explanations (see Figure \ref{fig:teaser}). Users can click through the argument tree to explore nested reasoning chains at their own pace.
    \item \textit{Socratic Chatbot}: Instead of showing errors directly, the chatbot asks targeted questions about specific arguments (see Figure \ref{fig:ui}). Only after users verbalize the issue does the system add a comment marker as a revision reminder. A progress indicator shows completion across all identified issues. In contrast to the Visual Feedback method, only the sentences directly relating to the current Socratic conversation are highlighted.
\end{itemize}
\end{itemize}

\subsection{Design}
Our design choices follow established principles from cognitive psychology and Human Computer Interaction to support active reasoning rather than passive acceptance of AI-generated content.

\subsubsection{Delayed feedback during writing.} Unlike real-time grammar checkers, Critical Inker provides no feedback while typing. Research in cognitive load theory demonstrates that immediate, fragmented feedback can increase extraneous cognitive load and interfere with the generative thinking required during initial composition \cite{fyfe2015}. While immediate feedback has been shown to reduce errors in procedural tasks, delayed feedback better supports metacognitive development and deeper understanding in complex learning contexts \cite{moreno2004}. 

% By withholding analysis until users complete their drafts or press the analyze button, we preserve the uninterrupted flow of ideation essential to the writing process.

\subsubsection{Verbalization requirement (Socratic mode).} Research on self-explanation demonstrates that learners who actively generate explanations to themselves achieve significantly deeper understanding and improved problem-solving performance compared to those who passively receive information \cite{danry2023don, chi1989self,chi1994eliciting}. The self-explanation effect occurs when students articulate their reasoning, which prompts them to identify gaps, generate inferences, and repair mental models \cite{chi1989self}. Based on prior work in Socratic questioning \cite{paul2008critical, danry2023don}, the Socratic chatbot is prompted to not give away the answer, but rather support the user to arrive at the identified logical errors themselves. Only when the user identifies the issue themselves will the Socratic chatbot create a comment marker reminding the user about what they want to change. This verbalization requirement transforms internal reflection into concrete action, ensuring engagement translates into text revision.

\subsubsection{Question grounding in argument structure.} Socratic questioning is most effective when it is systematic, disciplined, and focused on specific foundational concepts rather than generic prompts \cite{paul2008critical}. Effective Socratic questions probe specific claims, assumptions, and logical relationships through targeted inquiries that require students to justify their reasoning \cite{paul2008critical}. Rather than asking generic questions like "Is this claim supported?", Critical Inker's chatbot references specific logical relationships extracted from the argument graph: \textit{"You claim X because Y, but I am curious, how does Y actually support X?"} This specificity, enabled by LLM-based argument mining, helps users locate precise gaps in their reasoning and engage with the actual structure of their arguments to fix them.

\subsubsection{Progressive disclosure.} Progressive disclosure is an interaction design technique that reduces cognitive overload by revealing information sequentially rather than all at once \cite{nielsen2006progressive,spillers2010progressive}. Research shows that when users face too many simultaneous stimuli, their working memory capacity can be exceeded, leading to errors, missed information, and task abandonment \cite{sweller1988cognitive}. The Socratic chatbot addresses one argument at a time rather than overwhelming users with all issues simultaneously. The progress bar provides a sense of accomplishment while maintaining focus on individual reasoning chains, following established principles of staged disclosure \cite{carroll1984training}.

\subsection{Argument Extraction and Evaluation}
\label{sec:prompting}

Critical Inker uses a multi-stage prompting pipeline for both interaction modalities: (1) Structure Extraction, (2) Logical Evaluation, and (3) Socratic Intervention. We split structure extraction, evaluation, and Socratic dialogue into independent subtasks to separate the context allowing for different outputs.

\subsubsection{Task 1: Structure Extraction}
The first prompt (\ref{sec:prompt_1}) transforms the user's essay into a argumentation graph. LLMs struggle reliably return character indices or exact quotes. To solve this, we explicitly instruct the model to return ``atomic quotes'' for every claim or premise identified. Combined with fuzzy string-matching this worked well in our evaluations.

Furthermore, the prompt instructs the model to distinguish between \textit{independent reasons} (which support a claim on their own) and \textit{joined reasons} (which only function as support when combined). This distinction is encoded in the JSON output structure, where independent reasons are arrays of single items \texttt{[id, target]} and joined reasons are nested arrays \texttt{[[id1, id2], target]}. Using notation and language from literature about informal logic helped us to get more reliable outputs.

\subsubsection{Task 2: Logical Evaluation}
We observed that single-shot prompting (extracting structure and evaluating validity simultaneously) led to degraded performance. Instead we decided to iterate through each support relation found in Phase 1 and validate it individually using a dedicated Evaluation Prompt (\ref{sec:prompt_2}).

This prompt forces the model to engage in chain-of-thought reasoning before outputting a verdict. The JSON schema requires a \texttt{rationale} field (``Think through logical validity step by step assuming that the evidence is true'') before the \texttt{strength} field (``valid'' or ``invalid'').

\subsubsection{Task 3: Socratic Questioning}
For the chatbot modality, we designed a system prompt that strictly forbids direct correction (\ref{sec:prompt_4}). The instructions emphasize a ``reasoning assistant'' persona that ``interestedly inquires'' about the user's intent. 

\noindent The chatbot is provided with the full JSON analysis from Phase 2. It tracks the conversation state against a generated plan (\ref{sec:prompt_3}) ensuring it only addresses one logical flaw at a time. If a flaw is resolved by the user it will create a comment citing what the user intended to fix.

To ensure robustness, all model interactions enforce strict JSON schemas either through the model's native ``tool use'' capabilities (e.g., Anthropic's \texttt{json\_response} tool) or JSON-mode constraints. 

\section{Technical Validation}
To create helpful interactions, the system requires a low-latency and high-accuracy analysis of the argumentation of the text. Therefore we evaluated the prompts with different models to show the idea is feasible and inform our model choice.

\subsection{Methodology}
We evaluated the prompts against established human-labeled datasets on how well they understand the structure of the argumentation and if they are able to detect if an argument is valid.

\begin{itemize}
\item \textbf{Structure:} We used the \textit{Argument Annotated Essay v2} dataset (TU Darmstadt) \cite{Stab2017} to test the model's ability to correctly identify claims and premises.

\item \textbf{Validity:}We adapted the \textit{Stanford Natural Language Inference (SNLI)} dataset \cite{Bowman2015}, converting logical pairs (premise \& hypothesis) to test the validity.
\end{itemize}
For both datasets we choose a random sample size of 100 essays or logical pairs.

\subsection{Prompt Evaluation Results}

% We are looking for models that perform similarly to humans when it comes to understanding argumentation in essays. Additionally, we care about response time so that users of the tool do not have to wait.

For the argument structure extraction we evaluated the models on three key metrics. 
(1) \textbf{Main claim:} All tested models performed similarly with $\approx 90\%$ accuracy, indicating that current models successfully grasp the informal logic and writer intention. 
(2) \textbf{Relations:} The model achieved a relations overlap of $91.2\%$. While slightly lower than the top score ($92.9\%$). 
(3) \textbf{Latency:} Claude Sonnet 4.5 achieved a mean execution time of $6.58$s, approximately $12\%$ faster than GPT-4.1 ($7.48$s) and $\sim 5\times$ faster than Gemini Flash ($37.12$s).

%\textbf{Argument Structure Extraction}

%\begin{itemize}
%    \item \textbf{Main claim:} All tested models performed similar detecting the main claim with around $90\%$ accuracy. This is an indicator that all current models understand the informal logic behind argumentative essays and understand the intention of the writer.
%    \item \textbf{Relations:} The model achieved a relations overlap of $91.2\%$. While slightly lower than the top score ($92.9\%$), this difference falls within standard margins of variance, indicating that Sonnet 4.5 captures complex argument graphs with state-of-the-art fidelity while significantly optimizing for speed.
%    \item \textbf{Time:} Claude Sonnet 4.5 achieved a Mean Execution Time of $6.58s$, approximately $12\%$ faster than GPT-4o ($7.48s$) and $\sim5\times$ faster than Gemini Flash ($37.12s$) on this complex task.
%\end{itemize}

\begin{figure}[h]
  \centering
  \includegraphics[width=\linewidth]{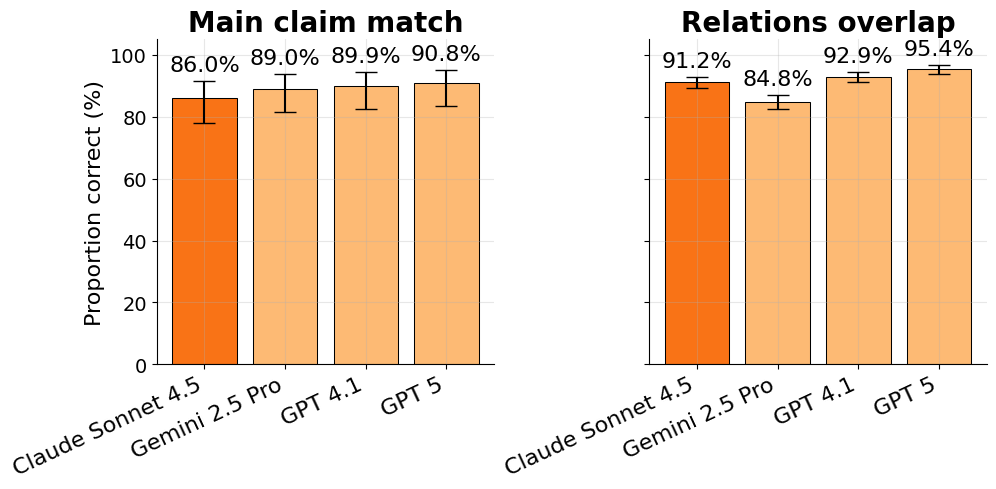}
  \caption{Structural accuracy across four LLMs compared to human ground truth from AAE v2 dataset.}
  \Description{}
  \label{fig:latency}
\end{figure}

For the validity check we established a robust baseline using Claude Sonnet 4.5. The model achieved a Validity Score of $87.0\%$ on the sample size. This performance is comparable to e.g., GPT-4.1 at $93\%$, suggesting the system is highly capable of minimizing false negatives in the critical feedback path, while staying in a reasonable and predictable time.

\begin{figure}[h]
  \centering
  \includegraphics[width=\linewidth]{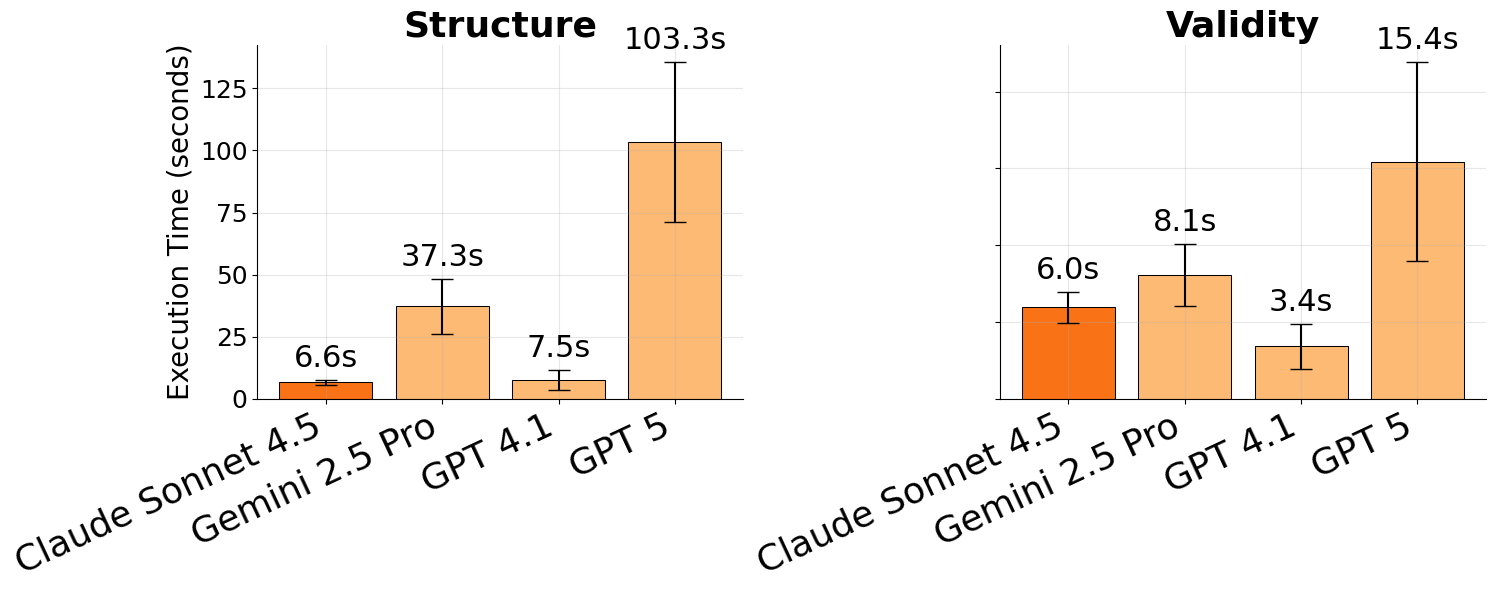}
  \caption{Latency comparison: Claude Sonnet 4.5 shows significantly lower variance ($\sigma=0.93s$) compared to GPT-4.1.}
  \label{fig:latency}
\end{figure}

\subsection{Preliminary Qualitative Results}
To gain early insights into user experiences with Critical Inker, we conducted a small-scale pilot study with 7 participants recruited online and randomly assigned to one of two conditions: \textit{Visual Feedback} (n=3), where logical flaws were directly highlighted with explanations, or \textit{Socratic Chatbot} (n=4), where the system guided users through targeted questions to identify issues themselves. Each participant wrote a short argumentative essay on a topic of their choice, received feedback from the assigned modality, and answered a question about their thinking process when using the tool. We analyzed interview transcripts using thematic analysis \cite{braun2006using,braun2019reflecting}.

In the \textit{Visual Feedback} condition, participants appreciated the clarity of the analysis. One participant mentioned that it ``helped identify logical weaknesses and guiding clearer and more precise revision,'' while another liked that errors ``were clearly shown to me in a way that made me understand how to remedy my own mistakes.'' However, one participant also mentioned a workflow friction, noting that they ``found it quite complex.''

In the \textit{Socratic Chatbot} condition, participants emphasized the value of active cognitive engagement. One participant noted that the tool ``let me think for myself instead of just writing it for me,'' while another stated it ``helped me to think about different aspects of the topic.'' However, one participant also mentioned friction, noting that ``it asked a question which was answered in the subsequent sentences,'' but concluded that ``discussing/arguing over it with the AI [...] encouraged me to flesh out my point a little more.''

\section{Discussion \& Conclusion}

In this paper, we present Critical Inker, a prototype in development that uses logical analysis and Socratic interactions to promote critical thinking while writing. The prototype introduces two novel interaction modalities: a Socratic chatbot that bases its questions on specific logical connections rather than general semantics and visual feedback that enables users to interactively explore the argument structure. Our technical evaluation confirms the feasibility of the LLM-based argument mining, achieving 91.2\% structural overlap with human annotators and 87\% validity accuracy with low latency ($\sigma=0.93$s). Preliminary qualitative feedback highlights a tension between efficiency and engagement. While the Visual Feedback condition offered clarity, participants noted it felt "complex," suggesting that exposing full logical structures may increase cognitive load. Conversely, the Socratic Chatbot indicated some reflection effects ("let me think for myself"), but occasionally introduced friction by asking questions the user felt were already addressed. This suggests that both the visual feedback and Socratic chatbot modalities may promote critical reflection but require further study.

%%
%% The acknowledgments section is defined using the "acks" environment.
\begin{acks}
This work was conducted at the Fluid Interfaces Group at the MIT Media Lab. Special thanks to Joanne Leong for foundational exploration in this area. Furthermore we want to thank Prof. Albrecht Schmidt for supporting the collaboration.
\end{acks}

% -------------------------------------------------------------------------
% REFERENCES
% -------------------------------------------------------------------------
\bibliographystyle{ACM-Reference-Format}
\bibliography{_references}

\appendix

\section{Prompts}
\label{sec:appendix_prompts}

To ensure reproducibility, we provide the exact system prompts used for the analysis pipeline.

\subsection{Prompt 1: Structure Extraction}
\label{sec:prompt_1}
This prompt is used to decompose the argumentation structure of the essay. \{\{ESSAY\_CONTENT\}\} will be replaced by the essay in the code.

\begin{lstlisting}
Analyze the argumentation structure of the following essay and output 
it in JSON format.

First identify and quote the main claim of the essay. This should be 
the author's core position.

Analysis:
1. Extract every distinct argumentative statement (reasons, premises, 
   evidence) as separate atomic quotes
2. Map which reasons support which claims
3. Distinguish direct support (statements that directly support the 
   main claim) from indirect support (statements that support other 
   supporting statements)
4. Identify joined reasons (work ONLY together) vs independent reasons
5. Continue tracing support chains until reaching axioms (unsupported 
   base claims)

JSON Format:
{
  claim: {
    "content": "author's core position in your words",
    "claim_quote": "exact quote of the main thesis",
    "support_relations": {
      "quotes": { 
        "1": "exact quote atomic reason in this case for claim_quote",
        "2": "exact quote atomic reason 2",
        ...
      },
      "relations": [ 
        # [key of quote(s) supporting claim/premise/evidence, 
           target claim/premise/evidence] where target is what gets 
           supported
        ...
      ]
    }
  }
}

KEY RULES:
- The authors main position should be clear from reading the main claim
- 0 always refers to claim_quote (the main thesis only - NOT sub-claims
  or premises)
- Quote character by character, preserving all errors, typos, 
  punctuation, case distinction, formatting and similar. Verify that 
  the quotes are exact.
- Ensure that independent and joint reasons are added correctly in 
  the json.
- Independent reasons (1->0; 2->0) are single arrays [1, 0], where 
  each reason only offers support for the conclusion without needing 
  any other premise.
- Joined reasons (2&3 -> 0) are arrays like [[2,3], 0], where reasons 
  (here 2 and 3) only work when both are present, i.e. one alone 
  provides insufficient support for the conclusion.
- Every quote must be atomic (one distinct claim)
- Claims & reasons can be shorter than a sentence
- Is there no argumentation set main claim to "" and keep quotes and 
  relations empty

ESSAY:
{{ESSAY_CONTENT}}
\end{lstlisting}

\subsection{Prompt 2: Validity Checking}
\label{sec:prompt_2}
This prompt determines whether the logic between a specific premise (evidence) and its target (claim) is valid.

\begin{lstlisting}
What is used to support this sentence "[content]".

Return your answer in the following json format:
{
    "claim":"[content]",
    "evidence":[evidence] # the premise that is used to support the claim. 
    This could be citations and/or another claim in the text.
    "evaluation":{
            "rationale":"Think through logical validity step by step 
            assuming that the evidence is true. Finally, report validity 
            strictly", # if references/citations are made to evidence 
            outside the text, assume that the reference/citation is true.
            "strength":"logic validity", # options ["valid", "invalid"]
            "rationale_short":"The main error to be communicated to the 
            user example: 'The evidence does not directly address... '" 
            # in simple understandable language
            "requirements": "What is necessary for the evidence to be 
            logically valid." # in short, simple and actionable language
            "label":"simple one-two words describing the logical flaw 
            (if any)" # if the argument is logically valid return "none".
            "label_long":"short definition of the label describing the 
            logical flaw (if any)" # if the argument is logically 
            valid return "none".
    }
}

Original Essay:
\end{lstlisting}

\subsection{Prompt 3: Plan for Socratic method}
\label{sec:prompt_3}

\begin{lstlisting}
Based on the argument analysis provided, create a step-by-step plan to fix all logical flaws in the text.
Each step should address one specific issue identified in the evaluation.

Focus on evaluations where strength is "invalid" and create actionable steps.
Order the steps from most critical to least critical issues.

Return your response in JSON format:
{
  "steps": [
    {
      "stepNumber": 1,
      "description": "Brief description of what needs to be fixed",
      "targetText": "The exact quote that has the issue",
      "issue": "What's wrong with it (from the evaluation)"
    }
  ]
}

If there are no flaws to fix, return an empty steps array.

Argument analysis:
\end{lstlisting}

\subsection{Prompt 4: Socratic assistant}
\label{sec:prompt_4}

\begin{lstlisting}
You are a reasoning assistant that helps people reasoning through logical flaws in their essays using the socratic method.
You are provided with an argument analysis with support relations between claims and evidence as.
The argument analysis also includes evaluations of the logical validity of relations between claims and evidence.

You guide the user using socratic questioning to help them realize their reasoning errors (logically invalid reasoning) and learn how to fix their errors.
The socratic method involves subtly directing the users intention towards a flaw in their reasoning without explicitly pointing out that it is a flaw.
You do this through multiple messages. Slowly direction more in each message.
Instead, interestedly inquires about the flaw asking what the user meant to say. And then ask questions to make them realize the flaw themselves.
After the user have realized the mistake you through questions help them correct the error in their text. Ask them if you should provide a suggestion for improvement.

Once the error is corrected, you move on to the next flaw. If there are none, you say that you have no feedback to provide & end the conversation.

You engage with the user conversationally as a dialogue.

**Output Format:**
Your response must be in the exact JSON format with the following structure:

{
  "messageToUser": Your conversational message to the user,
  "sentenceToUser": Exact quote of the sentence you want the user to think about,
  (optional) "suggestion": {
    "claim_quote": {"original": "exact quote of the main thesis", "suggestion": "suggestion for the main thesis"},
    "support_relations": [{"original": "exact quote atomic reason in this case for claim_quote", "suggestion": "suggestion for the atomic reason"}]
  }
}

KEY RULES:
- Focus on the main claim and support relations where strength is not "logically valid" and respect the "requirements" mentioned for improvement
- Answer in maximum 400 characters

Argument analysis results:
\end{lstlisting}

\end{document}